\documentclass{article}

\usepackage{PRIMEarxiv}

\usepackage{fancyhdr}
\usepackage[utf8]{inputenc} 
\usepackage[T1]{fontenc}    
\usepackage{hyperref}       
\usepackage{url}            
\usepackage{booktabs}       
\usepackage{amsfonts}       
\usepackage{nicefrac}       
\usepackage{microtype}      
\usepackage{xcolor}         

\usepackage{amsmath,amssymb,amsfonts}
\usepackage{makecell}
\usepackage{multirow}
\usepackage{graphicx,subcaption,lipsum}
\usepackage{algorithm}
\usepackage{algpseudocode}
\usepackage{verbatim}       
\usepackage{amsthm}

\title{Feasibility Study of CNNs and MLPs for Radiation Heat Transfer in 2-D Furnaces with Spectrally Participative Gases
}
\author{
  Axel TahmasebiMoradi, Vincent Ren, Benjamin Le-Creurer, Chetra Mang and Mouadh Yagoubi \\
  IRT SystemX \\
  2 Bd Thomas Gobert, 91120 Palaiseau, France \\
  \texttt{\{a.tahmasebimoradi, chetra.mang, mouadh.yagoubi\}@irt-systemx.fr} \\
  \texttt{\{vincent.ren, benjamin.le-creurer\}@airliquide.com} \\
}

\begin{document}
\maketitle

\begin{abstract}
Aiming to reduce the computational cost of numerical simulations, a convolutional neural network (CNN) and a multi-layer perceptron (MLP) are introduced to build a surrogate model to approximate radiative heat transfer solutions in a 2-D walled domain with participative gases. The originality of this work lays in the adaptation of the inputs of the problem (gas and wall properties) in order to fit with the CNN architecture, more commonly used for image processing.
Two precision datasets have been created with the classical solver, ICARUS2D, that uses the discrete transfer radiation method with the statistical narrow bands model.
The performance of the CNN architecture is compared to a more classical MLP architecture in terms of speed and accuracy. Thanks to Optuna, all results are obtained using the optimized hyper parameters networks. The results show a significant speedup with industrially acceptable relative errors compared to the classical solver for both architectures. Additionally, the CNN outperforms the MLP in terms of precision and is more robust and stable to changes in hyper-parameters.
A performance analysis on the dataset size of the samples have also been carried out to gain a deeper understanding of the model behavior. 
\end{abstract}


\section{Introduction}
With the introduction of massively parallel computers in the early 2000s, the representation of the solutions offered in computational fluid dynamics (CFD) has increased. Nevertheless, industry and academia are always working to minimize the CPU time for a given accuracy since it remains a limiting factor. It is well known that thermal radiation modeling in CFD requires a lot of memory and time. 
The choice of discretization techniques, mathematical model, and assumptions made about the behavior of the participating gases all have a significant impact on the numerical cost and accuracy of radiative heat transfer simulations. In order to solve the Radiative Transfer Equation (RTE), which requires both a spatial and an angular integration \cite{malalasekera2002calculation,modest2021radiative}, a large computational cost is involved. The gas behaves like a non-transparent medium, with a wavelength-dependent transmissivity coefficient, which presents an extra challenge.

A further discretization across the wavelengths is required when taking into account the spectral aspects. Thus, it is necessary to consider the spectral dependency of the walls and participating gases properties. Using the Statistical-Narrow-Band (SNB) model \cite{howell2020thermal}  with the Curtis-Godson (CG) approximation \cite{howell2020thermal,curtis1952statistical,godson1953evaluation} is one method for determining the transmissivity coefficient of each ray's element.

For most industrial applications, the computational cost of classical RTE solvers is prohibitive, which is their main drawback. Using artificial neural networks \cite{mishra2021physics,lagerquist2021using,chappell2022sunnynet,lu2022solving} and reduced order modeling techniques \cite{tencer2017accelerated,fagiano2016order}  to replace classical RTE solver is one method of cutting the computational cost. The non-intrusive, primarily geometry-dependent models in the previously mentioned works do not take into account the participative gases with spectral aspects in their mathematical modeling. 

This work examines how the parameters of the gas depend on the spectral wavelength. The RTE is solved using the Discrete Transfer Radiation Method (DTRM) for a domain with participative gases, and it is implemented in a FORTRAN solver known as ICARUS2D, which was co-developed in the 1990s by the EM2C laboratory \cite{EM2C} and Air Liquide \cite{airliquide}. In a 2D enclosed rectangular geometry, ICARUS2D returns the radiative heat fluxes on the walls after solving the radiative heat transfer for a mixture of participative gases. Since its results are thought to offer high-fidelity solutions, they will be used as the reference ones for training a metamodel. 

The current work aims to propose a computationally faster alternative to the classical RTE solver (ICARUS2D). Given that the RTE solution depends on the boundary and domain values, this is not an easy task. We will observe that both Multi-Layer Perceptrons (MLP) and Convolutional Neural Networks (CNN) can learn the behavior of radiative heat transfer problem for a 2-D furnace, and they are faster than the classical solver, ICARUS2D.  

\subsection{Main contribution}
As stated in the introduction, the computational cost of solving radiative transfer equation with spectral participative gases is high. To our best knowledge, there are not many works that have used machine-learning methods to replace RTE classical solver. Therefore, the main contributions of this work is to investigate the usage of CNN and MLP on a 2D dataset based on solutions of RTE with spectral participative gases. To make the CNN compatible with radiation problem, each type of input parameters (such as emissivity, temperature of the boundary and temperature of the domain) are considered as a unique input channel.

\section{Radiative Heat Transfer in an Industrial High-Temperature Furnace}
\subsection{Radiative Transfer Equation}
The steady-state radiative transfer equation for a spectral participative, non-scattering gas is  \cite{malalasekera2002calculation,modest2021radiative}  

\begin{equation}
\frac{\partial I_\nu (\vec{r},\vec{s})}{\partial s}=\kappa_\nu I_{b\nu} \left(T\left(\vec{r}\right)\right)-\kappa_\nu I_\nu \left(\vec{r},\vec{s}\right)
\end{equation}
where sub-script $\nu$ refers to the spectral dependency of the corresponding variables, $\vec{s}$ is the direction along the ray, $\vec{r}$ is the location in the spatial space, $I_\nu $ is the radiative intensity, $\kappa_\nu$  is the absorption coefficient, $T$ is the gas temperature, and $I_{b\nu}$  is the Planck’s function. For a diffusive wall (a surface that reflects diffusively) with a known temperature $T^0$, the boundary condition at a given point ${\vec{r}}^0$ is given by

\begin{equation}
\begin{split}
I_\nu^0\left({\vec{r}}^0,0\right)=I_\nu^0\left(0\right)=\varepsilon_v I_{b\nu} \left(T^0\right)+ \\
\frac{1-\varepsilon_v }{\pi}\int_{\vec{n}.{\vec{s}}^\prime<0}{I_\nu \left(s^\prime\right)}\left|\vec{n}.{\vec{s}}^\prime\right|d\Omega^\prime
\end{split}
\end{equation}

where 0 refers to a location on a boundary where the ray is drawn, $\vec{n}$ is the normal of the boundary, $\varepsilon_v$  is the spectral emissivity of the boundary, ${\vec{s}}^\prime$ is the direction along the incident/incoming ray and $\Omega^\prime$ is the solid angle corresponding to ${\vec{s}}^\prime$. By integrating over spectral bands, the hemispherical irradiation on a boundary surface can be written as:

\begin{equation}
H=\int_{0}^{\infty}H_\nu d\nu=\int_{\vec{n}.{\vec{s}}^\prime<0}\int_{0}^{\infty}{I_\nu \left(s^\prime\right)}d\nu\left|\vec{n}.{\vec{s}}^\prime\right|d\Omega^\prime
\end{equation}

\subsection{Statistical Narrow Band Model}
The SNB model with the Curtis-Godson (CG) approximation can be used to calculate the spectral absorption coefficients of participative gases for a given spectral band. It is outside the scope of this work to discuss the SNB-CG model; interested readers should consult \cite{howell2020thermal,curtis1952statistical,godson1953evaluation} for further information. Based on SNB-CG model for gases, the absorption coefficients are functions of temperature ($T$), wave number ($\nu$), total pressure ($p$), and molar fractions of participative gases ($x_g$).

\subsection{Discrete Transfer Radiation Method}
Proposed by Lockwood and Shah \cite{lockwood1981new}, the Discrete Transfer Radiation Method (DTRM) is used for its acceptable compromise between accuracy and computational cost. The DTRM solves radiative rays in a spatially and angularly discretized space domain. When the spectral aspects are considered, an additional discretization over the wavelengths has to be done. As mentioned earlier, this is the method used in the classical solver, ICARUS2D.

\subsection{2D Industrial Furnace}
\begin{figure}[h!]
    \centering
    \subfloat[]{
    \includegraphics[width=0.4\textwidth]{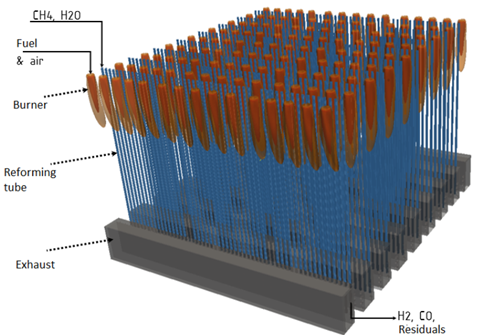}}
    \qquad
    \subfloat[]{
    \includegraphics[width=0.4\textwidth]{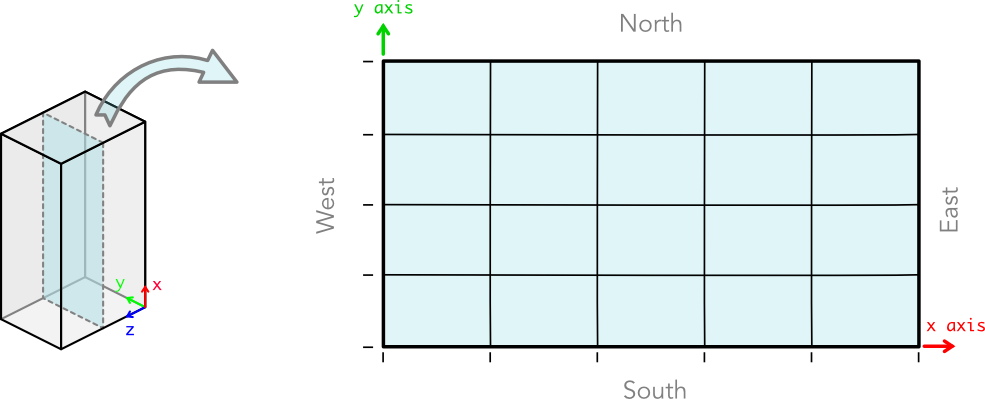}}
    \caption{schematics of an industrial furnace for methane reforming process, a) 3D and b) 2D}
    \label{fig:3D_2D_furnace}
\end{figure}

In reality, an industrial furnace is a 3D and not perfectly a rectangular domain. However, for the sake simplicity, we consider a 2D rectangular shape furnace. Moreover, CO$_2$, CO and H$_2$O gases are considered. The boundary and the domain are meshed in a Cartesian way. For this problem, the input parameters are emissivities ($\varepsilon$) and temperature ($T_0$) of the boundary, and the temperature ($T$) of the domain. On the other hand, the output is the hemispherical irradiation ($H$) for each boundary point. For $n_x$ and $n_y$ discretization points on the boundaries, we have $2(n_x + n_y)$ boundary points and $(n_xn_y)$ domain points. Schematics of 3D and 2D furnaces are shown in Fig. \ref{fig:3D_2D_furnace}

\section{Neural Networks}
In this section, we present briefly two well-known neural network architectures; 1- CNN, and 2- MLP. 

\subsection{MLP}
\begin{figure}[h!]
    \centering
    \subfloat[]{
    \includegraphics[width=0.48\textwidth]{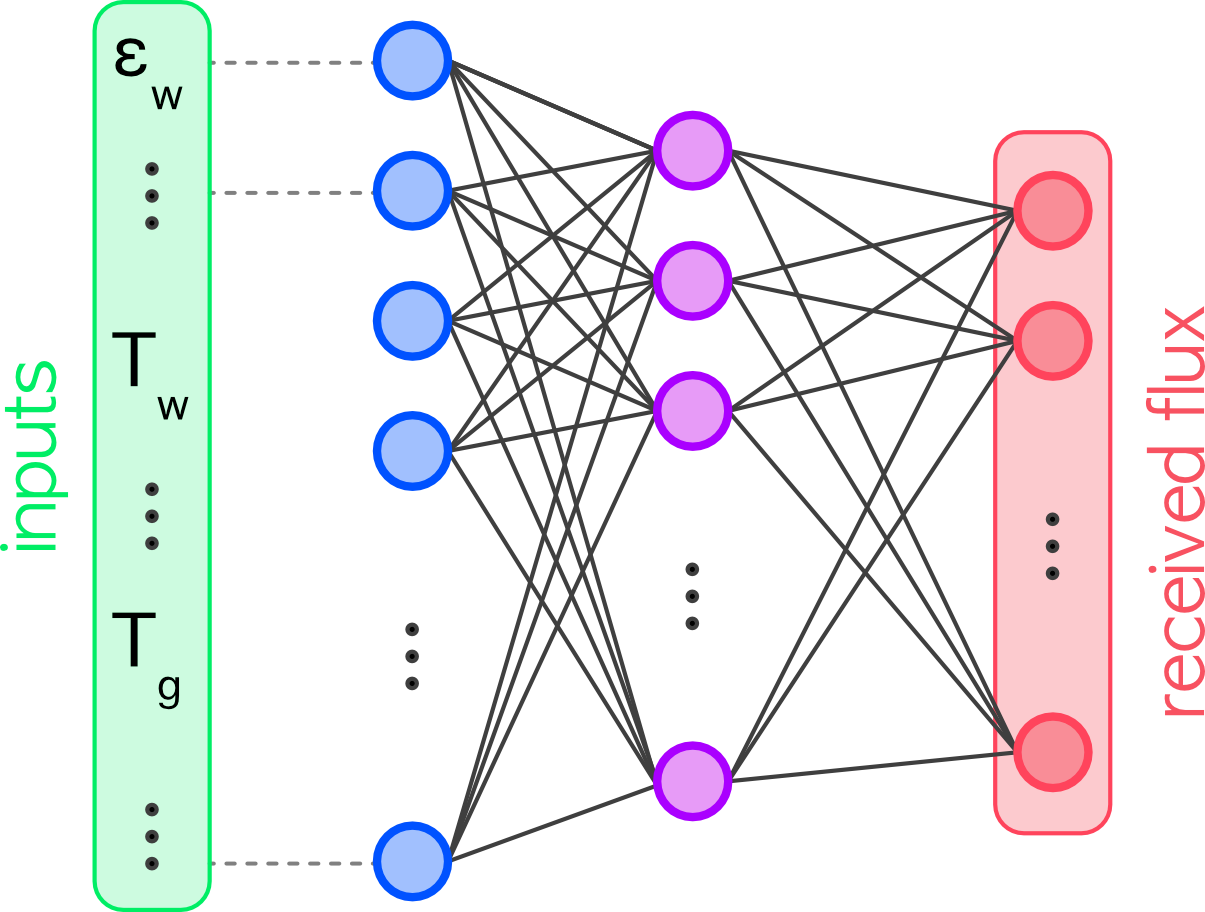}}
    \caption{MLP architecture}
    \label{fig:MLP_architecture}
\end{figure}

Multi-layer perceptrons are the most classical type of neural networks. For the present case, the input layer is composed of emissivities, boundary temperatures and domain temperatures arranged in the input vector. The output layer is the hemispherical irradiation on boundary points (see Fig~\ref{fig:MLP_architecture}). For this case, we consider the first boundary point to be the one on the very left of the south wall and the remaining boundary points are followed in a counter clock-wise manner. 

\subsection{CNN}
\begin{figure}[h!]
    \centering
    \subfloat[]{
    \includegraphics[width=0.17\textwidth]{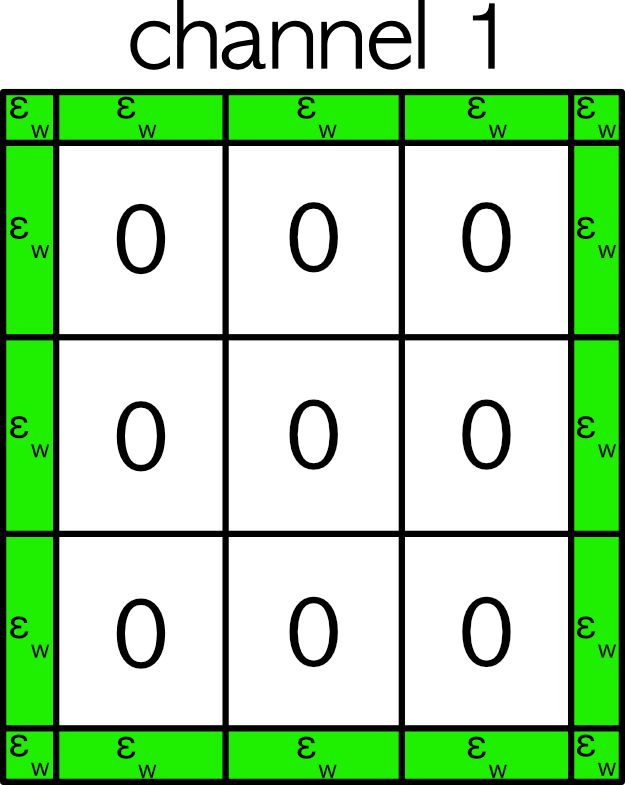}}
    \qquad
    \subfloat[]{
    \includegraphics[width=0.17\textwidth]{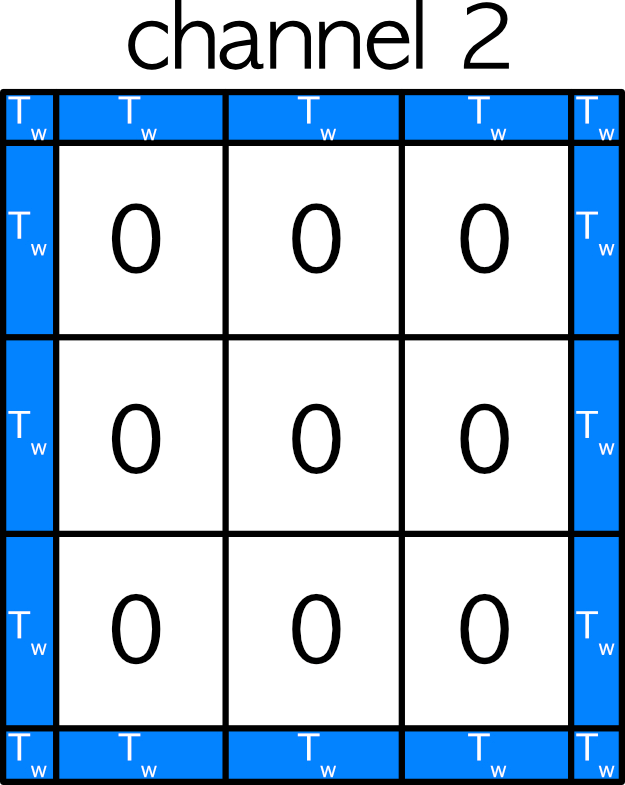}}
    \qquad
    \subfloat[]{
    \includegraphics[width=0.17\textwidth]{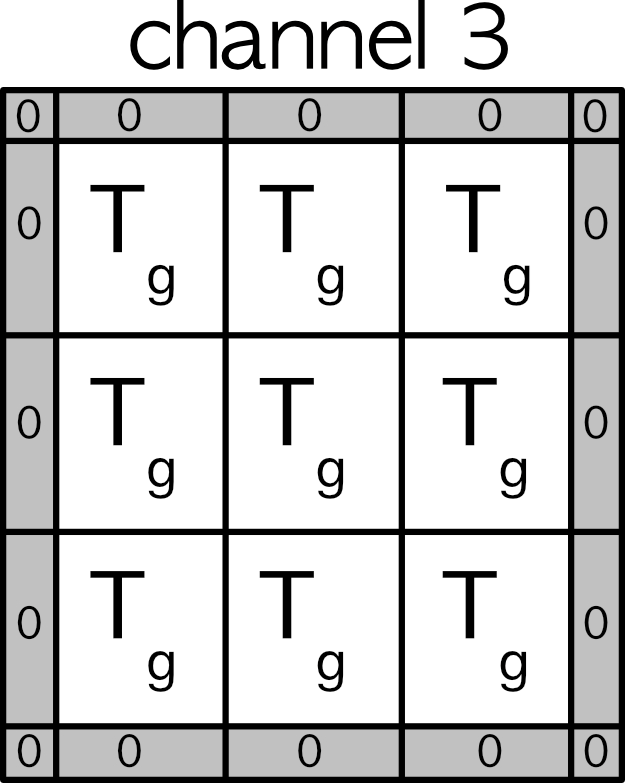}}
    \qquad
    \subfloat[]{
    \includegraphics[width=0.48\textwidth]{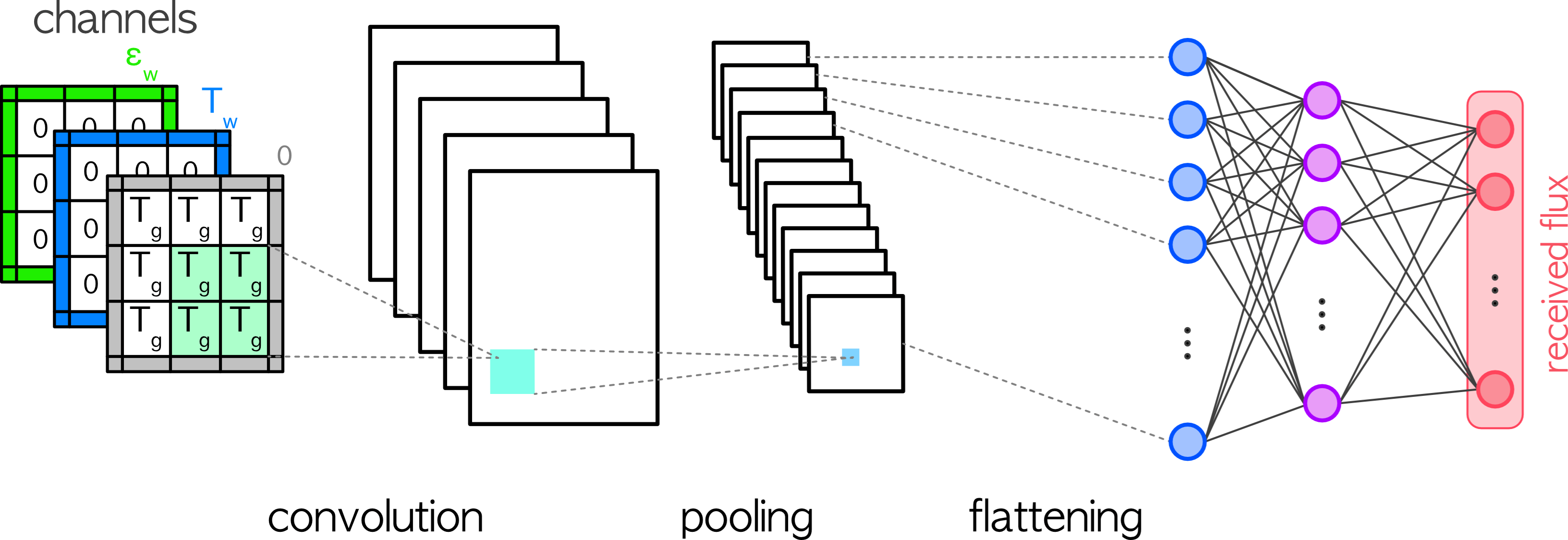}}
    \caption{CNN architecture}
    \label{fig:CNN_architecture}
\end{figure}

Convolutional neural networks are mostly used for image processing tasks. Since a 2D industrial furnace resemble a rectangular image, with some assumptions, a CNN can be used. We define the input matrix (image) as depicted on Figure~\ref{fig:CNN_architecture}.

As can be seen, each input parameter (emissivity, boundary temperature, domain temperature) is considered as a channel. We just set the values to zero for those that don't exist, e.g. wall temperature and emissivity in the domain, domain temperature on the wall. The output layer is the same as the one used with MLP architecture.

\section{Results and discussions}
The geometrical and physical properties of the 2D furnace are presented in Table \ref{table:table_2Ddomain_properties}.

\begin{table}[h!]
\centering
\begin{tabular}{|cc|ccc|c|}
\hline
\multicolumn{2}{|c|}{spatial discretization}    & \multicolumn{3}{c|}{lengths (m)}  & {angular discretization}        \\ \hline
\multicolumn{1}{|c|}{$n_x$} & \multicolumn{1}{c|}{$n_y$} &    \multicolumn{1}{c|}{$L_x$}&  \multicolumn{1}{c}{$L_y$} &   & \multicolumn{1}{c|}{\#rays}            \\ \hline
\multicolumn{1}{|c|}{120} & \multicolumn{1}{|c|}{20} & \multicolumn{1}{|c|}{12.0} & \multicolumn{1}{|c}{2.0} & &\multicolumn{1}{|c|}{32} \\ \hline
\multicolumn{2}{|c|}{molar concentration of gases}    & \multicolumn{3}{c|}{wave number ($cm^{-1}$)} &  {pressure ($atm$)}   \\ \hline
\multicolumn{1}{|c|}{$x_{CO_2}$} & \multicolumn{1}{c|}{$x_{H_2O}$} & \multicolumn{1}{c|}{$\nu_{min}$} & \multicolumn{1}{c|}{$\nu_{max}$} & \multicolumn{1}{c|}{$\Delta \nu$} & p                  \\ \hline
\multicolumn{1}{|c|}{0.1} & \multicolumn{1}{c|}{0.2} & \multicolumn{1}{c|}{150.0} & \multicolumn{1}{c|}{9300.0}  & 25.0 & 1.0   \\ \hline
\end{tabular}
\caption{Geometrical and physical properties of the 2D furnace }
\label{table:table_2Ddomain_properties}
\end{table}

In order to investigate the efficiency of neural networks for radiative heat transfer problem, we created two datasets using Latin-Hypercube Sampling (LHS) method: 

\begin{itemize}
  \item Dataset A: 1000 training samples with 300 test samples, and
  \item Dataset B: 3000 training samples with 900 test samples.
\end{itemize}

For the preliminary results, we used the dataset A. For all the simulations, the optimized hyper-parameters are found using Optuna. Based on the recent research on the minimum width of neural networks \cite{park2020minimum}, we chose the following values/ranges of parameters to be used in Optuna (Table \ref{table:hyper_parameters_ranges}),

\begin{table}[h]
\centering
\begin{tabular}{|c|c|c|c|}
\hline
\multicolumn{1}{|c|}{network type} & \multicolumn{1}{c|}{\# MLP layers} & \multicolumn{1}{c|}{\# nodes} & - \\ \hline
\multicolumn{1}{|c|}{MLP} & \multicolumn{1}{c|}{$[1,3]$} & \multicolumn{1}{c|}{$[1000,10000]$} & - \\ \hline
\multicolumn{1}{|c|}{CNN} & \multicolumn{1}{c|}{$[1,3]$} & \multicolumn{1}{c|}{$[20,1000]$} & - \\ \hline
{\# conv. layers} & \multicolumn{1}{c|}{\# filters}    & \multicolumn{1}{c|}{filter sizes} & \multicolumn{1}{c|}{pooling sizes}  \\ \hline
$[1,3]$ & \multicolumn{1}{c|}{$[3,27]$} &  \multicolumn{1}{c|}{$[1,6]$} & \multicolumn{1}{c|}{$[1,6]$}  \\ \hline
\end{tabular}
\caption{Minimum and maximum values of the hyper-parameters used in MLP and CNN }
\label{table:hyper_parameters_ranges}
\end{table}

 The other hyper-parameters are fixed and their values are shown in Table \ref{table:hyper_parameters_fixed}. 
The optimized hyper-parameters for both networks are presented in Table \ref{table:optimized_hyper_parameters}. 

 \begin{table}[h]
\setlength{\tabcolsep}{5pt}
 \centering
\begin{tabular}{|c|c|c|c|c|c|}
\hline
 parameter &  \multicolumn{1}{|c|}{ dropout }  & \multicolumn{1}{c|}{optimizer type}  & \multicolumn{1}{c|}{learning rate} & \multicolumn{1}{c|}{batch size}   & \multicolumn{1}{c|}{\# epochs}  \\ \hline
 value & \multicolumn{1}{|c|}{0.0} & \multicolumn{1}{|c|}{Adam} & \multicolumn{1}{c|}{$20000$}& \multicolumn{1}{c|}{$0.001$} &   \multicolumn{1}{c|}{$20000$}  \\ \hline
{loss function} & \multicolumn{1}{c|}{kernel regularizer}  & \multicolumn{1}{c|}{activation function} & \multicolumn{1}{c|}{stride} & \multicolumn{1}{c|}{conv.'s padding} & \multicolumn{1}{c|}{pooling type}  \\ \hline
MAE & \multicolumn{1}{c|}{$0.0011$} &\multicolumn{1}{c|}{elu} & \multicolumn{1}{c|}{$(1,1)$} & \multicolumn{1}{c|}{same} & \multicolumn{1}{c|}{Average Pooling2D}  \\ \hline
\end{tabular}
\caption{Fixed values of the hyper-parameters used in MLP and CNN }
\label{table:hyper_parameters_fixed}
\end{table}

\begin{table}[h]
\centering
\begin{tabular}{|c|c|c|c|}
\hline
\multicolumn{1}{|c|}{network type} & \multicolumn{1}{c|}{\# MLP layers} & \multicolumn{1}{c|}{\# nodes} & - \\ \hline
\multicolumn{1}{|c|}{MLP} & \multicolumn{1}{c|}{$1$} & \multicolumn{1}{c|}{$7405$} & - \\ \hline
\multicolumn{1}{|c|}{CNN} & \multicolumn{1}{c|}{$1$} & \multicolumn{1}{c|}{$724$} & - \\ \hline
{\# conv. layers} & \multicolumn{1}{c|}{\# filters}    & \multicolumn{1}{c|}{filter sizes} & \multicolumn{1}{c|}{pooling sizes}  \\ \hline
$1$ & \multicolumn{1}{c|}{$9$} &  \multicolumn{1}{c|}{$(2,3)$} & \multicolumn{1}{c|}{$(1,1)$}  \\ \hline
\end{tabular}
\caption{The optimized hyper-parameters used in MLP and CNN }
\label{table:optimized_hyper_parameters}
\end{table}

Using the obtained hyper-parameters, the mean and standard deviation (STD) values of relative errors calculated for all boundary points using both dataset A and B are presented in Table \ref{table:mean_std_all}. 


\begin{table}[h]
\centering
\begin{tabular}{|c|c|c|c|}
\hline
                  {dataset} & {network type} & {mean \%} & {standard deviation \%}  \\ 
                  \hline
\multirow{2}{*}{A} & MLP & 21.89 & 5.79    \\ \cline{2-4} 
                   & CNN & 9.07  & 1.84    \\ \hline
\multirow{2}{*}{B} & MLP & 19.67 & 5.77   \\ \cline{2-4} 
                   & CNN & 7.40 & 1.55   \\ \hline
\end{tabular}
\caption{Mean, standard deviation of the relative error of all wall points}
\label{table:mean_std_all}
\end{table}

By comparing these values, one can see that the CNN yields better results than the MLP, still the relative errors for both cases are high. One can also observe the effect of larger datasets on the training of both networks. For MLP, the larger dataset B improved the mean and STD values by $10.14\%$ and $0.34\%$, respectively. Similarly, for CNN, the mean and STD values are ameliorated by $18.41\%$ and $15.76\%$, respectively.

\begin{figure}[h]
    \centering
    \subfloat[]{
    \includegraphics[width=0.47\textwidth]{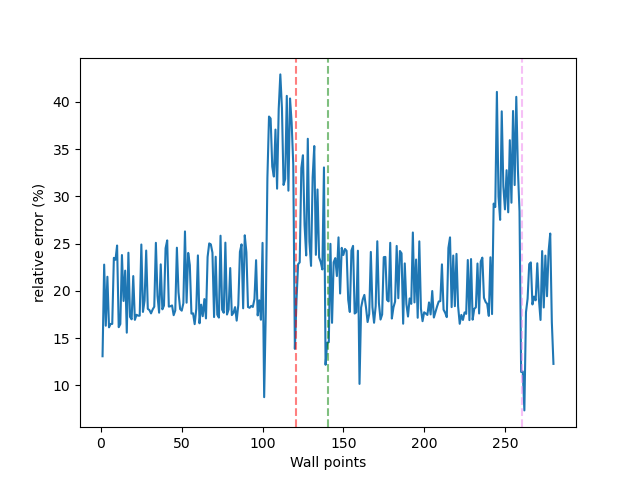}}
    \qquad
    \subfloat[]{
    \includegraphics[width=0.47\textwidth]{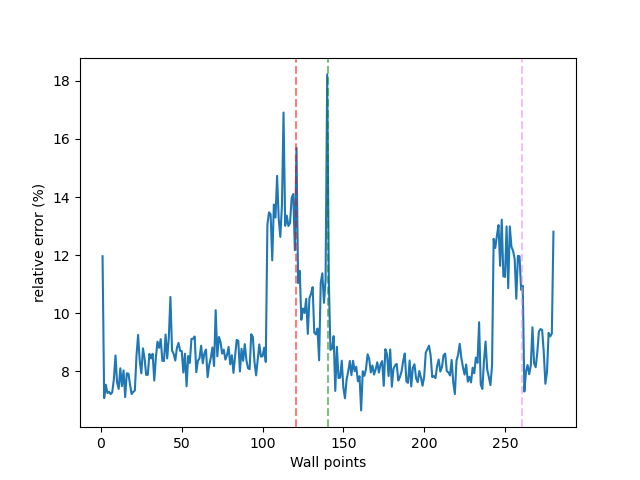}}
    \qquad
    \caption{relative error of boundary points, a) MLP and b) CNN }
    \label{fig:relative_errors}
\end{figure}

The relative errors for each network are shown in Fig. \ref{fig:relative_errors}. The points on the horizontal axis: 
\begin{itemize}
  \item from 0 to 120 (red line) belong to south wall,
  \item from 120 (red line) to 140 (green line) belong to east wall,
  \item from 140 (green line) to 260 (purple line) belong to north wall, and
  \item from 260 (purple line) to 280 belong to west wall.
\end{itemize}
Interestingly, we observe that for the north and south walls the results are better than those for the east and west walls. In other words, both networks have a tendency to predict better the hemispherical irradiation on the north and south walls. Also, we can see that for the south/north walls, the values of wall points at the extremities are not as good as the values of the points in the middle. 

All the simulations are carried out on a workstation with 32 gigabyte of RAM and a  Intel(R) Core(TM) i9-900 clocked at 3.10GHz CPU. The Python packages of TensorFlow 2.0 are used for creating/training/testing the neural networks. The training and inference times of both networks on both datasets are reported in Table \ref{table:train_inference_all}. One by one, each sample of test set was forward-sent in both network and then the average inference time were obtained. As expected, the training time of larger dataset B is more than the training time of dataset A. Furthermore, both training and inference times of CNN are less than those of MLP. Obviously, the difference comes from different hyper-parameters and architecture used in each networks (see Table \ref{table:optimized_hyper_parameters}). It is worth mentioning that by use of ICARUS2D (the classical solver), the simulation time for one sample is about 138 seconds. This means that we obtain an approximate x7000 speedup by using the networks. There are two reasons for this speedup: 

\begin{itemize}
  \item The RTE is solved using the traditional solver using DTRM. Not all of the DTRM steps are parallelized in the classical solver, and this approach is not fully parallelizable on multiple CPU/GPU cores. On the other hand, we can take advantage of a multi-cores workstation using TensorFlow 2.0 to train and use neural networks.  
  \item As the radiative intensity depends on the spectral band, the RTE should be solved for the whole domain one at a time for a spectral band. In other words,  RTE must be solved 367 times for a spectral band of 25 cm$^{-1}$, based on the wave number minimum (150 cm$^{-1}$) and maximum (9300 cm$^{-1}$) values. However, this for-loop on the wave number is avoided by use of the trained networks.     
\end{itemize}

\begin{table}[h]
\centering
\begin{tabular}{|c|c|c|c|}
\hline
                  {dataset} & {network type}& {training time (sec)} & {inference time (sec)}  \\ \hline
\multirow{2}{*}{A} & MLP & 11307 & 0.0210    \\ \cline{2-4} 
                   & CNN & 8296  & 0.0117    \\ \hline
\multirow{2}{*}{B} & MLP & 28794 & 0.0198   \\ \cline{2-4} 
                   & CNN & 22365 & 0.0122   \\ \hline
\end{tabular}
\caption{training and inference time of neural networks of all wall points}
\label{table:train_inference_all}
\end{table}

For the same dataset A, we use again MLP and CNN networks but this time, the output vector is only composed of one wall points. The new optimized hyper-parameters for these networks are presented in Tables \ref{table:optimized_hyper_parameters_south_wall} and \ref{table:optimized_hyper_parameters_east_wall} for south and east walls, respectively. 

\begin{table}[h]
\centering
\begin{tabular}{|c|c|c|c|}
\hline
\multicolumn{1}{|c|}{network type} & \multicolumn{1}{c|}{\# MLP layers} & \multicolumn{1}{c|}{\# nodes} & - \\ \hline
\multicolumn{1}{|c|}{MLP} & \multicolumn{1}{c|}{$1$} & \multicolumn{1}{c|}{$9302$} & - \\ \hline
\multicolumn{1}{|c|}{CNN} & \multicolumn{1}{c|}{$1$} & \multicolumn{1}{c|}{$217$} & - \\ \hline
{\# conv. layers} & \multicolumn{1}{c|}{\# filters}    & \multicolumn{1}{c|}{filter sizes} & \multicolumn{1}{c|}{pooling sizes}  \\ \hline
$1$ & \multicolumn{1}{c|}{$9$} &  \multicolumn{1}{c|}{$(2,3)$} & \multicolumn{1}{c|}{$(1,1)$}  \\ \hline
\end{tabular}
\caption{The optimized hyper-parameters for south wall used in MLP and CNN }
\label{table:optimized_hyper_parameters_south_wall}
\end{table}

\begin{table}[h]
\centering
\begin{tabular}{|c|c|c|c|}
\hline
\multicolumn{1}{|c|}{network type} & \multicolumn{1}{c|}{\# MLP layers} & \multicolumn{1}{c|}{\# nodes} & - \\ \hline
\multicolumn{1}{|c|}{MLP} & \multicolumn{1}{c|}{$1$} & \multicolumn{1}{c|}{$4426$} & - \\ \hline
\multicolumn{1}{|c|}{CNN} & \multicolumn{1}{c|}{$1$} & \multicolumn{1}{c|}{$168$} & - \\ \hline
{\# conv. layers} & \multicolumn{1}{c|}{\# filters}    & \multicolumn{1}{c|}{filter sizes} & \multicolumn{1}{c|}{pooling sizes}  \\ \hline
$1$ & \multicolumn{1}{c|}{$9$} &  \multicolumn{1}{c|}{$(2,3)$} & \multicolumn{1}{c|}{$(1,1)$}  \\ \hline
\end{tabular}
\caption{The optimized hyper-parameters for east wall used in MLP and CNN }
\label{table:optimized_hyper_parameters_east_wall}
\end{table}

Having trained the networks, the mean and standard deviation values of relative errors calculated for the south and east wall points are reported in Tables \ref{table:mean_std_south} and \ref{table:mean_std_east}, respectively. Unexpectedly, we can see that using the larger dataset B doesn't improve or even worsen the accuracy of the MLP networks for south and east walls. We ran the training for a few times and we observed the same finding. One explanation can be that for the larger dataset B, one has to find again new hyper-parameters for each network. This finding reveals that the CNNs are more robust and stable to the hyper-parameters changes than the MLPs.   

\begin{table}[h!]
\centering
\begin{tabular}{|c|c|c|c|}
\hline
                  {dataset} & {network type} & {mean \%} & {standard deviation \%}  \\ 
                  \hline
\multirow{2}{*}{A} & MLP & 9.77  & 7.76   \\ \cline{2-4} 
                   & CNN & 7.48  & 1.61  \\ \hline
\multirow{2}{*}{B} & MLP & 12.07 & 8.82   \\ \cline{2-4} 
                   & CNN & 5.57  & 1.28  \\ \hline
\end{tabular}
\caption{Mean and standard deviation of the relative error of south wall points}
\label{table:mean_std_south}
\end{table}

\begin{table}[h!]
\centering
\begin{tabular}{|c|c|c|c|}
\hline
                  {dataset} & {network type} & {mean \%} & {standard deviation \%}  \\ 
                  \hline
\multirow{2}{*}{A} & MLP & 2.90 & 1.06   \\ \cline{2-4} 
                   & CNN & 3.25 & 1.59  \\ \hline
\multirow{2}{*}{B} & MLP & 2.98 & 0.50  \\ \cline{2-4} 
                   & CNN & 2.58 & 1.01    \\ \hline
\end{tabular}
\caption{Mean and standard deviation of the relative error of east wall points}
\label{table:mean_std_east}
\end{table}

\begin{figure}[h!]
    \centering
    \subfloat[]{
    \includegraphics[width=0.47\textwidth]{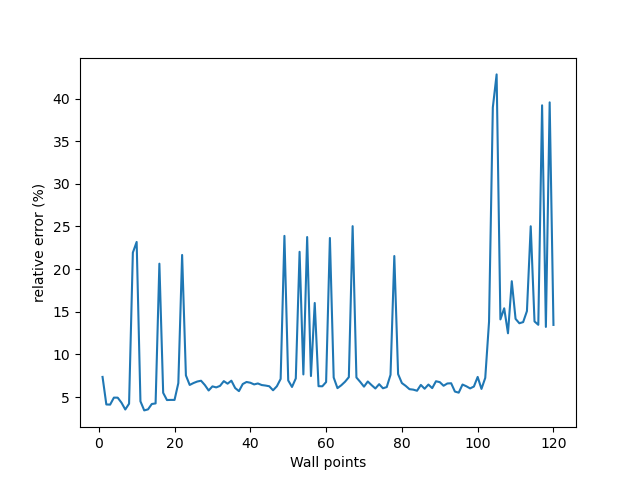}}
    \qquad
    \subfloat[]{
    \includegraphics[width=0.47\textwidth]{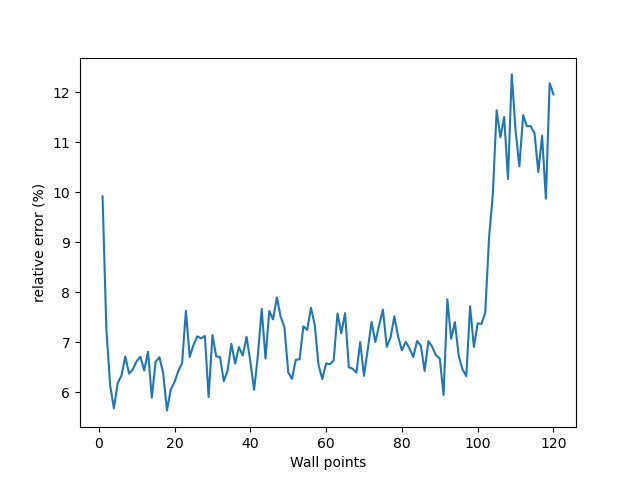}}
    \qquad
    \caption{relative error of south wall points, a) MLP and b) CNN }
    \label{fig:relative_errors_southwall}
\end{figure}

\begin{figure}[h!]
    \centering
    \subfloat[]{
    \includegraphics[width=0.47\textwidth]{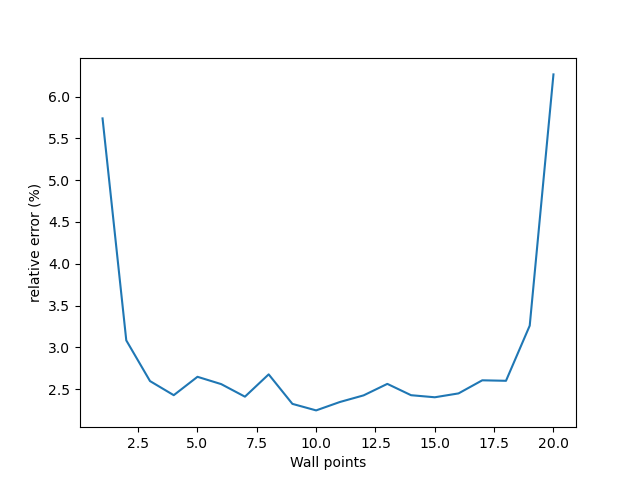}}
    \qquad
    \subfloat[]{
    \includegraphics[width=0.47\textwidth]{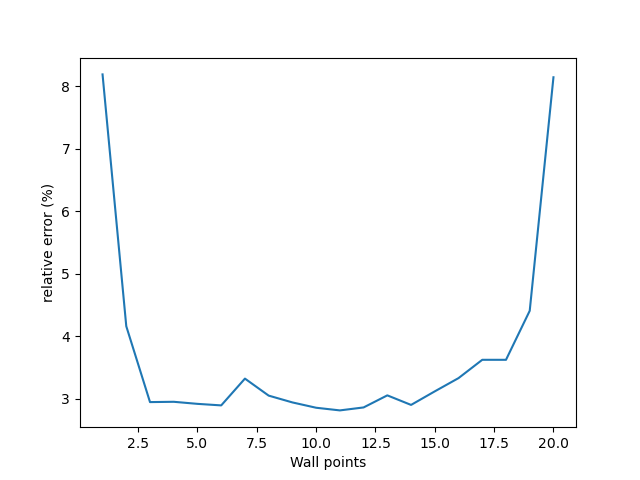}}
    \qquad
    \caption{relative error of east wall points, a) MLP and b) CNN }
    \label{fig:relative_errors_eastwall}
\end{figure}

The relative errors for south wall points are shown in Fig. \ref{fig:relative_errors_southwall}, and for east wall points are shown in Fig. \ref{fig:relative_errors_eastwall}.

 These results are much better in comparison of those shown in Table \ref{table:mean_std_all} for the same wall points. As can be seen, by considering only a wall, both networks have a better tendency to predict the hemispherical irradiation. It is worth noting that the same results obtained almost for north and west walls in comparison to the south and east walls, respectively. 

\begin{table}[h!]
\centering
\begin{tabular}{|c|c|c|c|}
\hline
                  {dataset} & {network type}& {training time (sec)} & {inference time (sec)}  \\ \hline
\multirow{2}{*}{A} & MLP & 13734  & 0.0231   \\ \cline{2-4} 
                   & CNN & 5006  & 0.0051  \\ \hline
\multirow{2}{*}{B} & MLP & 34800 & 0.0227   \\ \cline{2-4} 
                   & CNN & 13572  & 0.0046  \\ \hline
\end{tabular}
\caption{training and inference time of neural networks of south wall points}
\label{table:train_inference_south}
\end{table}

\begin{table}[h!]
\centering
\begin{tabular}{|c|c|c|c|}
\hline
                  {dataset} & {network type}& {training time (sec)} & {inference time (sec)}  \\ \hline
\multirow{2}{*}{A} & MLP & 6677 & 0.0122   \\ \cline{2-4} 
                   & CNN & 4290 & 0.0036  \\ \hline
\multirow{2}{*}{B} & MLP & 16388 & 0.0121  \\ \cline{2-4} 
                   & CNN & 12282 & 0.0034    \\ \hline
\end{tabular}
\caption{training and inference time of neural networks of east wall points}
\label{table:train_inference_east}
\end{table}

The training and inference times of the networks on both datasets are reported in Table \ref{table:train_inference_south} and \ref{table:train_inference_east} for south and east wall points, respectively. As before, the training time of larger dataset B is more than the training time of dataset A. Also, both training and inference times of CNNs are less than those of MLPs. Obviously, the difference comes from different hyper-parameters and architecture used in each networks (see Table \ref{table:optimized_hyper_parameters}).


\section{Summary}

In this work, a CNN and a MLP are introduced to build a surrogate model to approximate radiative heat transfer solutions in a 2-D walled domain with participative gases, with the goal of lowering the computational cost of numerical simulations. The novel aspect of this work is how the problem's inputs, i.e. gas and wall properties, were used to work with the CNN architecture. Using the classical solver called ICARUS2D, two high precision datasets were generated, and then used to train and test the networks. 

The results showed that both MLP and CNN can indeed learn the solution of RTE. Compared to MLP, CNN performed better in terms of accuracy and speedup. For the cases when the output of both networks corresponded to all the boundary points, the mean and standard deviation of the relative error of CNN was $9.07\%$ and $1.84\%$, respectively. Also, the speedup was about x11000. We also saw that larger dataset improved the accuracy of both networks. However, the improvement was bigger for CNN network than MLP network ($18\%$ v.s. $10\%$). Then, we chose the output vector to correspond to only one boundary wall. Having kept the same hyper-parameters, we observed that the new reduced-in-size output vector for only one boundary wall enhanced the accuracy of CNNs but not the MLPs. This means that the CNN outperformed the MLP in terms of robustness and stability against hyper-parameter changes.

In conclusion, both MLP and CNN can capture and learn the radiative heat transfer. These networks can replace, with a trade-off, the classical solvers when a lot of different simulations are needed. However, this requires to have/generate a decent number of samples, and this is a computationally costly procedure. For cases where a few simulations are only needed, it is better to use the classical solvers due to their high fidelity solutions with a lower computational cost since there is no need for generation of a high fidelity dataset.

\section*{Acknowledgements}
The authors also thank the industrial partners for their supports in the framework of the HSA project \cite{irtsysx}. This project is as well supported by the French government's aid in the framework of PIA (Programme d'Investissement d'Avenir) for Institut de Recherche Technologique SystemX.

\bibliographystyle{unsrt} 
\bibliography{references}

\begin{thebibliography}{10}

\bibitem{malalasekera2002calculation}
Weeratunge Malalasekera, Hendrik~K Versteeg, Jonathan~C Henson, and JC~Jones.
\newblock Calculation of radiative heat transfer in combustion systems.
\newblock {\em Clean Air}, 3(1):113--143, 2002.

\bibitem{modest2021radiative}
Michael~F Modest and Sandip Mazumder.
\newblock {\em Radiative heat transfer}.
\newblock Academic press, 2021.

\bibitem{howell2020thermal}
John~R Howell, M~Pinar Meng{\"u}{\c{c}}, Kyle Daun, and Robert Siegel.
\newblock {\em Thermal radiation heat transfer}.
\newblock CRC press, 2020.

\bibitem{curtis1952statistical}
AR~Curtis.
\newblock A statistical model for watervapour absorption.
\newblock {\em QJ Roy. Met. Soc.}, 78:639--640, 1952.

\bibitem{godson1953evaluation}
WL~Godson.
\newblock The evaluation of infra-red radiative fluxes due to atmospheric water
  vapour.
\newblock {\em Quarterly Journal of the Royal Meteorological Society},
  79(341):367--379, 1953.

\bibitem{mishra2021physics}
Siddhartha Mishra and Roberto Molinaro.
\newblock Physics informed neural networks for simulating radiative transfer.
\newblock {\em Journal of Quantitative Spectroscopy and Radiative Transfer},
  270:107705, 2021.

\bibitem{lagerquist2021using}
Ryan Lagerquist, David Turner, Imme Ebert-Uphoff, Jebb Stewart, and Venita
  Hagerty.
\newblock Using deep learning to emulate and accelerate a radiative transfer
  model.
\newblock {\em Journal of Atmospheric and Oceanic Technology},
  38(10):1673--1696, 2021.

\bibitem{chappell2022sunnynet}
Bruce~A Chappell and Tiago~MD Pereira.
\newblock Sunnynet: A neural network approach to 3d non-lte radiative transfer.
\newblock {\em Astronomy \& Astrophysics}, 658:A182, 2022.

\bibitem{lu2022solving}
Yulong Lu, Li~Wang, and Wuzhe Xu.
\newblock Solving multiscale steady radiative transfer equation using neural
  networks with uniform stability.
\newblock {\em Research in the Mathematical Sciences}, 9(3):45, 2022.

\bibitem{tencer2017accelerated}
John Tencer, Kevin Carlberg, Marvin Larsen, and Roy Hogan.
\newblock Accelerated solution of discrete ordinates approximation to the
  boltzmann transport equation for a gray absorbing--emitting medium via model
  reduction.
\newblock {\em Journal of Heat Transfer}, 139(12):122701, 2017.

\bibitem{fagiano2016order}
Lorenzo Fagiano and Rudolf Gati.
\newblock On the order reduction of the radiative heat transfer model for the
  simulation of plasma arcs in switchgear devices.
\newblock {\em Journal of Quantitative Spectroscopy and Radiative Transfer},
  169:58--78, 2016.

\bibitem{EM2C}
EM2C.
\newblock Available at \url{http:// em2c.centralesupelec.fr}.

\bibitem{airliquide}
Airliquide.
\newblock Available at \url{http:// www.airliquide.com}.

\bibitem{lockwood1981new}
FC~Lockwood and NG~Shah.
\newblock A new radiation solution method for incorporation in general
  combustion prediction procedures.
\newblock In {\em Symposium (international) on combustion}, volume~18, pages
  1405--1414. Elsevier, 1981.

\bibitem{park2020minimum}
Sejun Park, Chulhee Yun, Jaeho Lee, and Jinwoo Shin.
\newblock Minimum width for universal approximation.
\newblock {\em arXiv preprint arXiv:2006.08859}, 2020.

\bibitem{irtsysx}
Irt systemx.
\newblock \url{https://www.irt-systemx.fr/projets/hsa/}.
\newblock Accessed: 2024-05-20.

\end{thebibliography}

\end{document}